\documentclass[final,5p,times]{elsarticle} 
\usepackage{epsfig}  
\usepackage{amssymb,amsmath}
\usepackage{tabularx}
\usepackage[table,xcdraw, dvipsnames]{xcolor}
\usepackage{lineno}
\usepackage{natbib}
\usepackage[greek,english]{babel}
\usepackage{subcaption}
\usepackage{float}
\usepackage[colorlinks=true,linkcolor=blue,urlcolor=blue]{hyperref}
\usepackage{ulem}
\usepackage{hyperref}
\begin{document}
\journal{Ecological Modelling}
\begin{frontmatter} 

\title{From female choice to social structure: Modeling harem formation in camelids}
\date{\today}

\author[tomas]{Tomás Ignacio González\corref{cor1}}
\ead{tomignaciogon@gmail.com}
\cortext[cor1]{Corresponding author}

\author[guillermo]{Guillermo Abramson}
\ead{abramson@cab.cnea.gov.ar}

\author[fabiana]{María Fabiana Laguna}
\ead{lagunaf@cab.cnea.gov.ar}

\address[tomas]{Statistical and Interdisciplinary Physics Division, Centro Atómico Bariloche (CNEA), CONICET and Postgraduate Program in Biology, Universidad Nacional del Comahue. 
R8402AGP Bariloche, Argentina}

\address[guillermo]{Statistical and Interdisciplinary Physics Division, Centro Atómico Bariloche (CNEA), CONICET and Instituto Balseiro (Universidad Nacional de Cuyo). R8402AGP Bariloche, Argentina.}

\address[fabiana]{Statistical and Interdisciplinary Physics Division, Centro Atómico Bariloche (CNEA), CONICET and Universidad Nacional de Río Negro. R8402AGP Bariloche, Argentina.}

\begin{abstract}
Herbivorous wild species constantly strive to optimize the trade-off between energy and nutrient intake and predation risk during foraging. This has led to the selection of several evolutionary traits---such as diet, habitat selection, and behavior---which are simultaneously shaped by the spatio-temporal variability of the habitat. Among camelid species, polygyny is a prevalent behavioral strategy that encompasses both mating and foraging activities. This group-level behavior has multiple interacting dimensions, contributing to an interesting ecological and evolutionary complexity. 
We developed an individual-based stochastic model in which camelid females transition between different familial groups in response to their environmental conditions, aiming to maximize individual fitness.
Our results indicate that the behavioral strategy of individual females can shape, by itself, emergent population-level properties, including group size and fitness distribution. Furthermore, these properties are modulated, in a non-additive manner, by other factors such as population density, sex ratio and system heterogeneity. 

\end{abstract}


\date{\today}
\end{frontmatter}

\section{Introduction}

Foraging behavior has traditionally been regarded as a major subject of interest in the study of different species, based on the assumption that this behavior has been optimized through natural selection over the course of the natural history of each species within their particular habitat \citep{pyke1977optimal,sih1980optimal,stephens1986foraging}. Both abiotic and biotic external factors can affect individual fitness and decision-making processes \citep{sih1980optimal,puig2008habitat}.  
Natural selection shapes phenotypic traits within populations regarding these factors, with the ultimate goal of maximizing individual survival and reproductive success. In animal species, the phenotypic aspects subject to selection are diverse, including traits such as diet, habitat selection, and behavior. When considered optimal, these traits are expected to maximize the benefit-to-cost relation of individual actions \cite{krebs1984optimization, werner1981optimal, comportamiento_vicuna}.

In wild ungulate species, foraging decisions are generally governed by the trade-off between the maximization of energy and nutrient intake and the minimization of predation risk. However, this general view must be contextualized within the seasonal, climatic and spatial variability of habitats, which are particularly relevant in arid and semi-arid environments \citep{puig2008habitat, wurstten2014habitat,smith2019habitat}.
In South America, the two most dominant and representative ungulate species are members of the Camelidae family: the guanaco (\textit{Lama guanicoe}) and the vicuña (\textit{Vicugna vicugna}) \citep{wurstten2014habitat,donadio2010evaluating}.

Among these species, group living is the norm, consisting both in a polygynous mating system  (characterized by one male, multiple females and their offspring) and a foraging strategy with multiple benefits: enhanced predator detection, reduced individual predation risk, and a reduction of the time allocated to vigilance during foraging \citep{lucherini1996aggressive,iranzo2018predator}.
However, both of these behaviors are shaped by a multitude of complex, interacting factors. 

As a foraging strategy, group movement patterns and cohesion are influenced by habitat structure and seasonality, visibility, predation risk, and food availability and selection \citep{ puig2008habitat, smith2019habitat, iranzo2018predator}. 

The characteristics of the mating system actually respond to these ecological and behavioral factors, particularly given that males must defend sizable territories to attract and retain females. Moreover, a comprehensive understanding of this social dynamics requires consideration of the differing interests of males and females, which are not exactly the same. Male reproductive success typically increases with harem size, whereas the opposite holds for females, as evidenced by a decrease in the average number of calves per female, as observed in Bonacic et al. (2002). This is a very important aspect that represents an initial trade-off in the system, in which females must assess the benefits of belonging to a larger group and whether it maximizes their fitness, as part of a decision-making process involving reproduction and survival. Although harem members generally benefit from reduced vigilance demands and increased foraging time, compared to solitary individuals, this trade-off does not apply to the dominant male, who devotes a greater proportion of time to vigilance and territorial defense \citep{vila1994time, bonacic2002density, arzamendia2006habitat, puig2007distribucion, cassini2009sociality}. 

Another important factor of camelid social structure is the presence of bachelor groups: non-breeding, non-territorial units, usually composed of young males. Several studies have shown that these groups have their own internal dynamics, distinct from those of the familial groups. Bachelor groups tend to have a more flexible and fluid composition, frequently undergoing fusion and fission events. These dynamics are closely related to the development of territorial and aggressive behavior among males, often expressed through locomotive play. Notably, more aggressive males tend to separate from their group earlier, and may subsequently establish territories and recruit females \citep{lucherini1996aggressive, cassini2009sociality, vila1995spacing}.

However, bachelor males can also exert coercive pressure on solitary or harem-associated females through sexual harassment, which is not strictly restricted to the breeding season. Solitary males frequently pursue and attempt to seize females, potentially causing harm and influencing female movement between territories. In the case of familial females, if the resident male is unable to provide adequate protection, harassment by bachelor males can destabilize the harem structure \citep{cassini2009sociality, clutton1992mate}.

This theoretical background provides a good foundation for the understanding of the various trade-offs and environmental and social factors that a camelid female must evaluate during her decision-making process. However, such an understanding of individual behavior does not necessarily allow us to predict the group-level properties and distributions. 
Complex systems are defined as those which components and subsystems with  non-linear interconnections that are difficult to anticipate, manage, or deduce, due to the emergence of unexpected properties product of these interactions. Within  this framework, the aggregation of individual decisions made by independent females may give rise to emergent processes and population-level features \citep{breckling2005emergent,johnson2006emergent, sumpter2006principles}.

The detection and study of emergent properties in complex systems present a significant challenge given the limitations of traditional scientific methods. In response to this, mathematical modeling and computational approaches have emerged as powerful tools for connecting  theoretical ecological principles to the structure and dynamics of complex natural ecosystems \citep{schweber2000complex,xiao2015emergent, van2022ecological}.

The objective of this study is to develop and implement a individual-based stochastic model inspired by the social structure of wild camelids, with a focus on female choice. In this model, individuals make decisions regarding the familial group they join, based on the optimization of their specific \textit{payoff}. We then analyze how these multiple individual-level processes scale up to generate population-level patterns.

\section{Model description}
Our system consists of a constant population (without  reproduction or mortality events), comprising fixed numbers of females ($F_T$) and males ($M_T$). During the organization process of this population, females constantly go through a decision-making process, with the opportunity to choose a new group or male to associate with. Meanwhile, males can play two possible roles, one as a familiar male with associated females, providing them with an amount of resource, the other one as a bachelor male (with no females) which increases the pressure over the females of all the sytem through harassment. The females's decision-making processs will be influenced by both social and ecological factors, and the sum of each particular decision will define the status and role of the males, which only act as passive individuals. This simplification does not reflect the natural system, where males exhibit very active social behavior. However, by “disabling” male actions other than harassment, we aim to isolate and better understand the role of female decision-making in shaping population structure. For our model, we adopt an assumption similar to that of Sibly (1983) \cite{sibly1983optimal}, which suggests that individuals (in our case, females) behave as if they knew which group size would maximize their fitness under the current environmental conditions, and will switch groups if doing so is advantageous.

\subsection{Female payoff definition}
First, we formulate an equation to estimate the female's payoff, $Q$, considering multiple factors:
\begin{equation}
    Q(F_i,B)_i = \frac{\sigma_i R_i}{m+cF_i} -\frac{H + \beta B}{1 + F_i}.
 \label{eq:payoff}   
\end{equation}
This equation represents the net gain of a female if she were located in harem $i$, which currently contains $F_i$ females, along with the presence of $B$ bachelors in the system. The first term reflects the benefit of belonging to the group led by male $i$, who controls a territory with $R_i$ food resources. These resources are protected and preserved with an efficiency $\sigma_i$. This benefit is balanced by the denominator, defined as the subsistence cost of the harem: the sum of the male's subsistence cost, $m$, and the subsistence cost of each female ($c$), multiplied by the number of females $F_i$. 
As a first approximation, $c$ could be considered similar to $m$, but this relationship is not trivial. Males and females perform different tasks, which may result in distinct energetic demands for each sex \citep{cassini2009sociality, lung2007influence, marino2012indirect}.
We define the product of $R_i$, the quality of the territory, and $\sigma_i$, the quality of the male, as the \textit{group's intrinsic quality}.
The resources available in a male’s territory must be shared among all associated females, meaning that the portion received by each female decreases as group size increases. This can be related to patterns observed in wild populations, where the number of calves per female declines with group size, possibly as a consequence of reduced individual access to resources \citep{ bonacic2002density}. While that field study suggests a linear relationship, we opted for a non-linear term with a similar monotonic behavior, which we considered more appropriate based on ecological and evolutionary reasoning, as well as methodological concerns regarding the field data used in that study.

The second term of Eq.~(\ref{eq:payoff}) accounts for the natural costs that the female must face. $H$ represents the mean field cost of the environment (synthesizing hazards as predation, climatic events, inter-specific competition, etc), while $\beta$ accounts for the additional cost associated with bachelors harassment. All these costs are divided by the size of the familiar group---which is ($F_i + 1$) if joined---reflecting the idea that a larger group of females may dilute the individual risks, while smaller groups are more vulnerable. This formulation is supported by field observations showing that the time each individual spends on vigilance decreases non-linearly with group size \citep{marino2008vigilance}. Additionally, grouping may serve as a strategy against cold temperatures by reducing heat loss through huddling \citep{hudson1985bioenergetics, de1998daily}.

We can simplify Eq.~(\ref{eq:payoff}) by normalizing the first term with the male subsistence cost, $m$. This redefines the resource as $R'_i=R_i/m$, and the females' cost as $c'=c/m$. 

To reduce the clutter of the notation, we shall call them just $R$ and $c$, dropping the primes, but keeping in mind that they are defined per unit of the male subsistence cost. Equation~(\ref{eq:payoff}) becomes:
 \begin{equation}
    Q(F_i,B)_i =\frac{\sigma_i R_i}{1+cF_i} -\frac{H + \beta B}{1 + F_i}
    \label{eq:2}
\end{equation}

\subsection{Individual-based model and simulations}
As noted above, the size of the system is determined by the total number of individuals of each sex, $M_T$ and $F_T$. Each male (whether a bachelor or holding a harem) constitutes a possible group that each female can join. Thus, the state of the system at time $t$ can be represented by the vector:
\[
S_t = (n_0, n_1, n_2, n_3,......,n_{M_T}),
\]
where $n_i$ denotes the number of females in the $i$-th harem,  with the exception of $n_0$, which represents the number of single females. Since the population size is constant, the total number of females across all groups always equals $F_T$:

\[
F_T=\sum^{M_T}_{i=0} n_i(t) ~\forall t.
\]

We simulate the system using an iterative process in which each step corresponds to a Monte Carlo step (MCS), with no association to any specific unit of real time. At each step, each female evaluates every male $i$ using Eq.~(\ref{eq:2}), assuming she would be the only new member joining his harem—i.e., she computes $Q(F_i + 1, B)_i$, with $i = 1, \dots, M_T$—even though all other females are simultaneously making the same assumption.

\begin{figure}[h]
\centering
\includegraphics[width=\columnwidth]{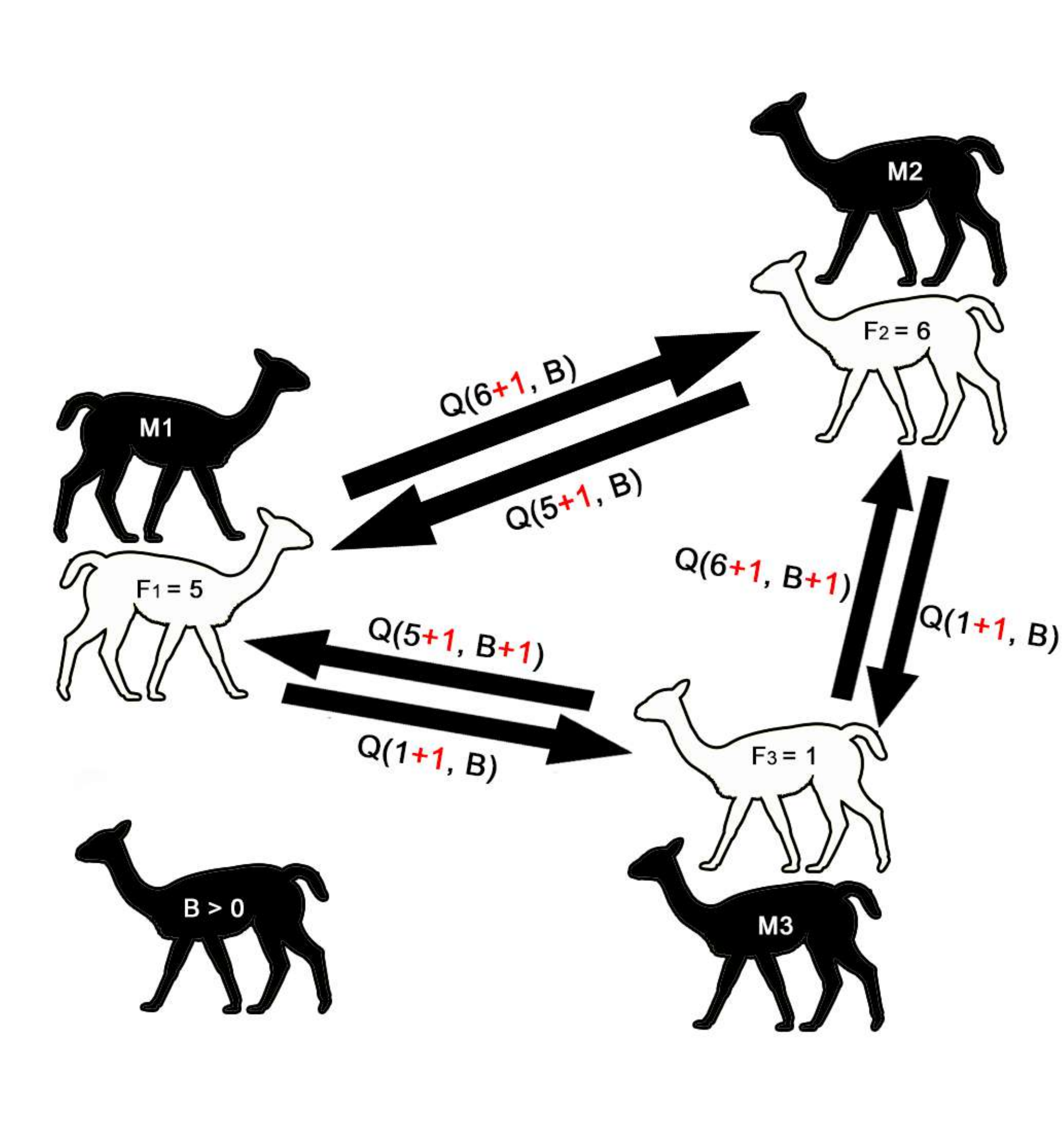}
\caption{Schematic representation of the payoff calculation network used to define each female's  probability distribution to switch group. In this example, the total number of females is $F_T$ = 12, with three existing harems and $B$ bachelors. Each female evaluates the potential payoff for each group based on her current position and the possible changes in the bachelors population.}
\label{fig:fig0}
\end{figure}

This results in a potential payoff distribution that will determine a probability distribution, based on which a stochastic process determines the female’s choice to join a harem. On this distribution, the probability of a male/harem to be chosen will be given by the relation between that group's potential payoff and the total potential payoff offer ($Q_T$).
But since $Q:\mathbb{Z}^2 \to \mathbb{R}$, the payoffs can be negative, which makes them unsuitable for a direct probability calculation. To overcome this issue, while preserving the overall shape of the distribution, we add the  absolute value of the minimum payoff in the current payoff distribution ($Q_{min}$) to each term, thereby shifting all values and ensuring that the least preferred male receives a probability of zero. We define, then, the probability distribution:
\begin{equation}
P =
\frac{1}{Q_T}
\begin{bmatrix}
Q(F_i+1,B+\Delta B)_1+|Q_{min}| \\
Q(F_i+1,B+\Delta B)_2+|Q_{min}| \\
Q(F_i+1,B+\Delta B)_3+|Q_{min}| \\
\cdots \\
Q(F_i+1,B+\Delta B)_{M_T}+|Q_{min}|
\end{bmatrix} = 
\begin{bmatrix}
p_1 \\
p_2 \\
p_3 \\
\cdots \\
p_{M_T}
\end{bmatrix},
\label{eq:prob}
\end{equation}
where the normalization is the total payoff offered by the males:
\[
Q_T=M_T|Q_{min}| + \sum^{M_T}_{i=1} Q(F_i+1,B+\Delta B)_i , 
\]
where $\Delta B$ represents the variation in the number of bachelors in the system that occurs when a female chooses a particular group, as illustrated in Fig.~\ref{fig:fig0}.

The initial state consists entirely of single individuals of both sexes, which also implies that $B = M_T$ bachelors are initially present in the system. This yields the initial state vector:
\[
S_0 = (F_T, 0, 0, 0,......,0).
\]
In order to advance to $t=1$, each female evaluates each bachelor male based on $Q(1,M_T-1)_i$, assuming she will be the only female in a new harem. These values define the initial probability distribution $P$, which is the same for all females at the initial state. 
Then, each female joins a group according to a multinomial process:
\[
S_{1} \in \text{Multinom}(S_0, P).
\]
In subsequent iterations, females will perform their own estimation (Eq.~(\ref{eq:2})) of the current state $S_t$, and perform their independent choices to switch group based on the corresponding  $P$ defined by Eq.~(\ref{eq:prob}).

It is important to notice that, once the females are distributed into harems, the probability distribution will vary between females in different harems, as each may be in a different situation.
For example, if a female from a large harem moves to another, she may only need to consider the difference in the number of females in the new harem and the group's intrinsic quality, $\sigma_i R_i$. 
However, if the only female in a harem of size 1 leaves it, the male will begin to act as a bachelor, increasing the negative costs to consider in the payoff calculation (see Fig.~\ref{fig:fig0}).

\subsection{Group switch penalty}
Even after disregarding active male behavior, females may still face costs associated with switching from one group to another. To represent this, we introduce a penalty term $\eta$ into the payoff distribution, applied prior to calculating the probability distribution $P$. If a female currently belongs to group  $z$, the modified payoff values are:

\[
Q'_i =
\begin{cases}
  Q(F_i,B)_i, & \text{if } i = z, \\
  Q(F_i,B)_i - \eta,   & \text{if } i \neq z.
\end{cases}
\]
The total normalized payoff $Q'_T$ and the minimum payoff $Q'_{min}$ are then redefined to include this penalty. Consequently, the resulting probability distribution $P'$ reflects a reduced likelihood that females switch to a different familial group, consistent with the idea of a social or energetic cost of relocation.

\section{Stochastic simulations}

In this section, we present the main results of our model simulations, focusing on how the sex ratio ($M_T/F_T$) and group intrinsic quality affect ($\sigma_iR_i$) population composition. We compare outcomes between a homogeneous system (where all groups have the same intrinsic quality) and a heterogeneous system (where group qualities vary). For this initial analysis, we fix all other parameters, $\beta = 1$, $H = 5$ and $c = 1$, which are sufficient to capture a representative range of system behaviors. However, the patterns and behaviors observed in the simulations remained qualitatively consistent across a broader range of parameter values ($\beta \in [0.1, 5]$, $c \in [0.1, 5]$, and $H \in [5, 100]$).

It should be emphasized that these ranges of the parameters do not correspond to any field measures and are purely theoretical, as one of our interests was to explore different scenarios of ecological relevance defined by different combinations of the parameters.

\subsection{Females distribution}

\begin{figure}[h]
\centering
\includegraphics[width=\columnwidth]{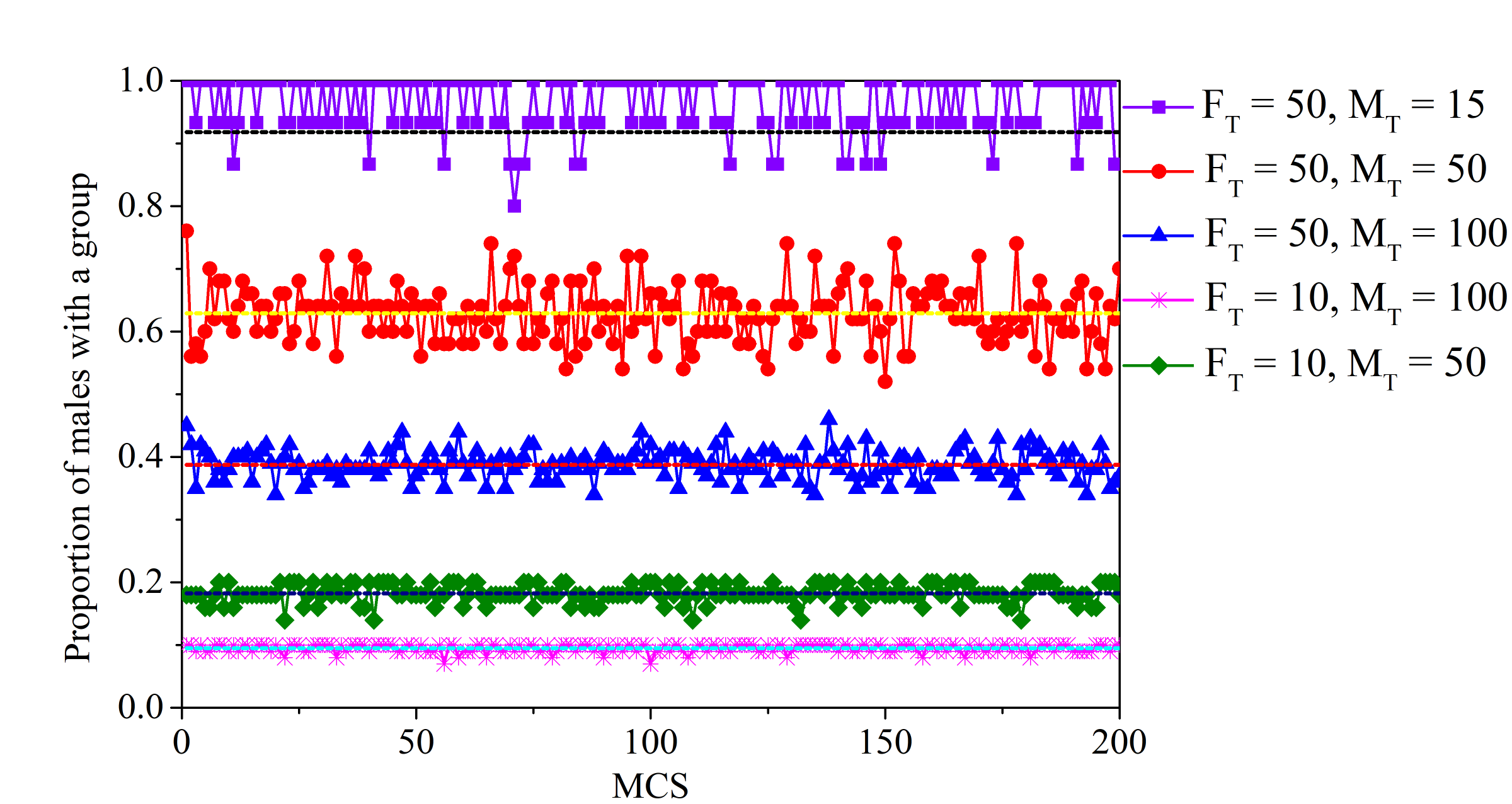}
\caption{Proportion of males with a group (harem) over time (MCS) for different combinations of $M_t$ and $F_t$, as indicated in the figure. In all cases, the parameters are fixed as follows:  $R = 20$, $H = 5$, $c = 1$, $\beta = 1$ and $\sigma = 1$. The respective averages are shown in a contrasting color.}
\label{fig:fig1}
\end{figure}

A first look at the evolution of this system suggests that the number of harems remains relatively stable over time. In Fig.~\ref{fig:fig1}, we plot the proportion of males with a group (referred to as 'familiar males') as a function of time; since each of these males represents a harem, this proportion effectively reflects the number of harems throughout the simulation. Additionally, the figure shows that the amplitude of fluctuations around this steady state varies with the total number of males ($M_T$) and females ($F_T$) in the system. Even when the main part of our simulations start at an initial condition $S_0$, as defined in the previous section, we also performed an additional set of simulations using different initial conditions. These variations had no effect on the steady state observed in Fig.~\ref{fig:fig1}.

\begin{figure}[h]
\centering
\includegraphics[width=\columnwidth]{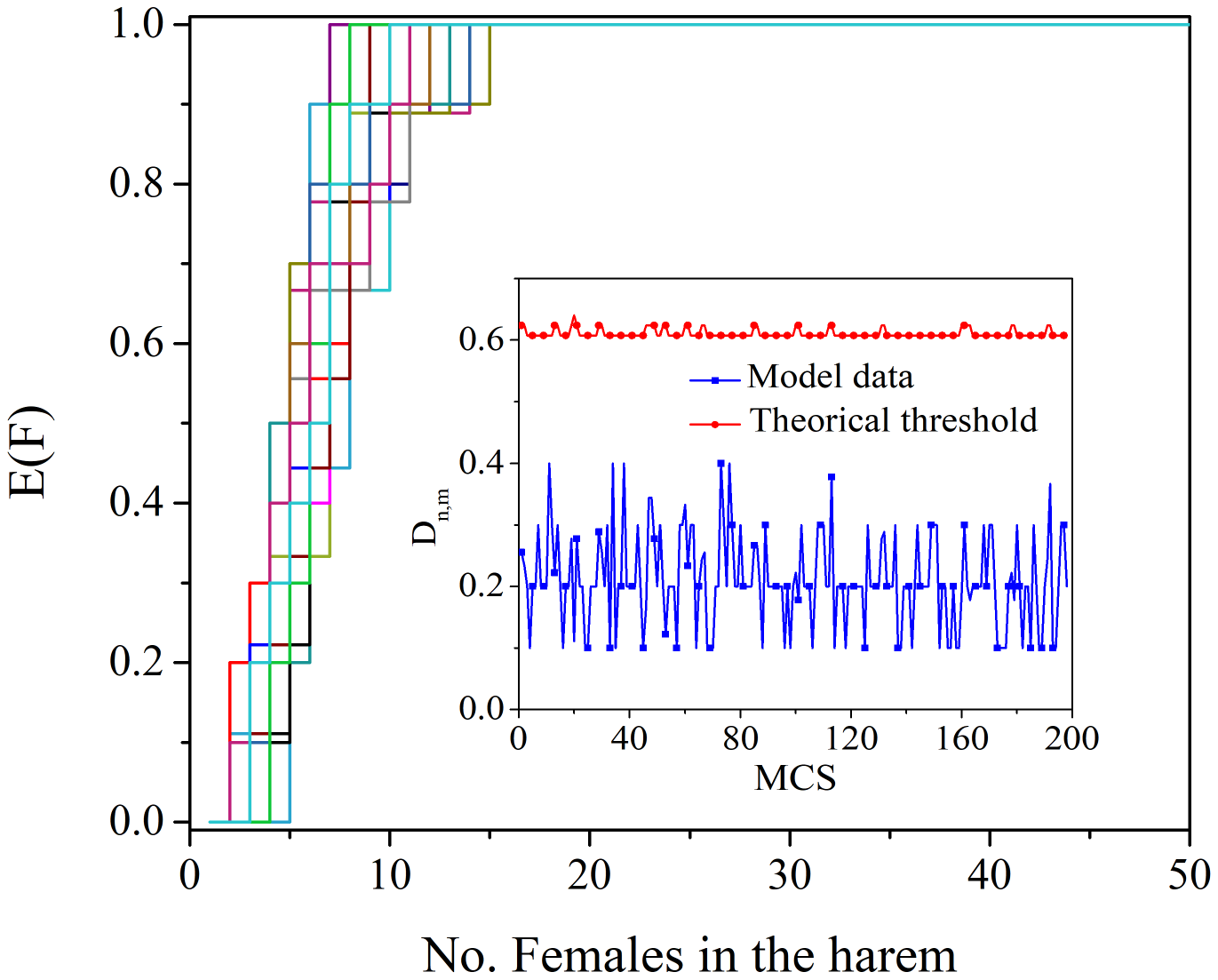}
\caption{Empirical cumulative density functions at different times $t$, calculated from the states vectors $S_t$. The inset shows the Kolmogorov–Smirnov distance between consecutive state vectors over time, compared to the theoretical threshold $D_c$.}
\label{fig:figK}
\end{figure}

To better characterize the system, it is necessary to know not only the number of harems but also the distribution of females within them, which is also a fluctuating quantity. We analyze the stability of the female's distribution through the visualization and comparison of the empirical cumulative distribution function (eCDF), which represents the fraction of groups that contain a number of females less than or equal to a given size $F$, at  time $t$:
\[
E_{t,n}(F) = \frac{n_{F_i \leq F}}n,
\]

where $n_{F_i \leq F}$ denotes the number of groups containing $F$ or fewer females, and $n$ is the total number of groups in the system.
In the main panel of Fig.~\ref{fig:figK}, we show a representative example of this function calculated at every time of the evolution of the system, in the stationary state. We observe that all the curves stay close to each other, indicating that the distribution of females remains similar across different stages ($S_t$) of the simulation.

To quantify this statement about the observations made in Fig.~\ref{fig:figK}, we calculate the Kolmogorov-Smirnov (K-S) distance, a measure of the difference between the eCDF function for the vector $S_t$ and the next iteration $S_{t+1}$. This is defined as the maximum distance between the two curves:
\[
D_{n,m} = \max (E_{t,n}(F) - E_{t+1,m}(F) ),
\]
where $E(F)$ are the eCDF functions for the states $S_t$ and $S_{t+1}$, and $n$ and $m$ are, respectively, the number of harems at times $t$ and $t+1$. 

According to statistical theory, we can define a threshold that indicates when the K-S distance between two state vectors is large enough to consider that they exhibit a significant difference in their distribution. This threshold is defined as:
\[
D_{c} = 1.36\, \sqrt{\frac{n+m}{nm}},
\]
where the value $1.36$ corresponds to a level of significance $\alpha=0.05$ \citep{harter1970selected,pratt1981kolmogorov}.

As shown in the inset of Fig.~\ref{fig:figK}, the K-S distance between consecutive states remains consistently below the theoretical threshold, supporting the conclusion that the system maintains a stable distribution of females over time. Although the figure presents a single representative case, this behavior was consistently observed across all simulations.

\subsection*{Homogeneous systems}
We define a homogeneous system as one where all territories are equally rich ($R_i = R, \forall i$) and all males are equally effective ($\sigma_i = \sigma, \forall i$).

\begin{figure}[h]
\centering
\includegraphics[width=\columnwidth]{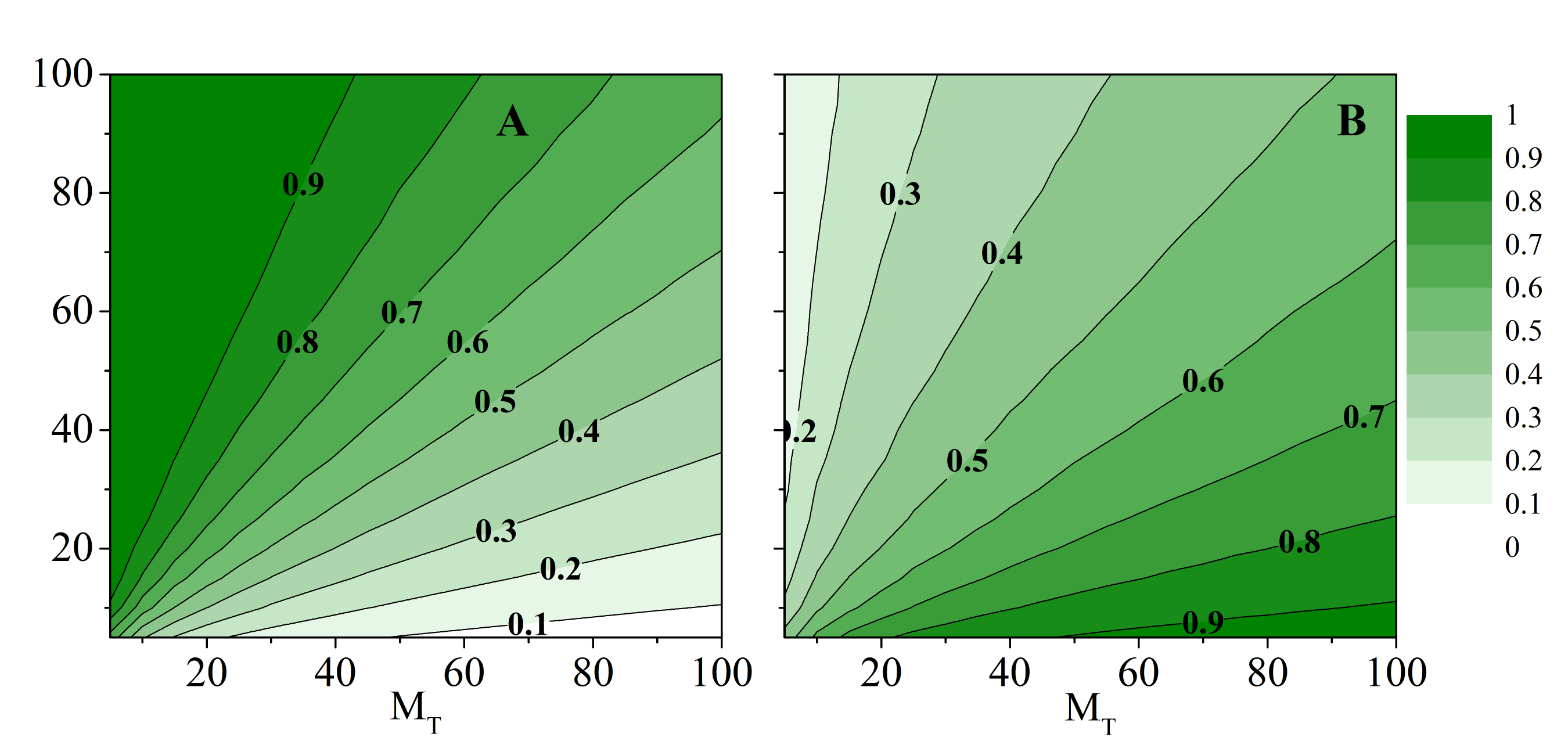}
\caption{(A) Heat map of the fraction of males with an harem, for different combinations of number of males ($M_t$) and females ($F_t$) in the system. (B) Heat map of the Global Gini index of the females distribution, also for different combinations of $M_t$ and $F_t$. Each pixel is the average of 20 realizations. Besides, $R = 20$, $H = 5$, $c = 1$, $\beta = 1$, $\sigma = 1$.}
\label{fig:fig2}
\end{figure}

A useful way to analyze the behavior of the system for different combinations of the parameters $F_T$ and $M_T$ is to represent the temporal average of the fraction of males with a harem (see Fig.~\ref{fig:fig2} A). We observe that this number increases as the ratio $F_T/M_T$ increases, as expected.

A similar pattern, but in the opposite direction, can be observed in the inequality of the distribution of females in the system. We characterize this using a Gini index---commonly used in economic sciences---that measures the inequality of the distribution of wealth. Values closer to 0 indicate a more uniform distribution, while values closer to 1 mean a greater inequality \citep{Farris2010}. The standard definition of the Gini index, adapted to our context, is given by:
 \begin{equation}
    G(t) =\frac{1}{2M_T F_T} \sum_{i,j}\left| n_i(t)-n_j(t) \right|,
    \label{eq:g2}
\end{equation}
where $n_i$ denotes the number of females  associated with the $i$-th male, as previously defined.  (In economic contexts, this variable typically represents individual wealth.)

We can observe in Fig.~\ref{fig:fig2} B that regions with a large $M_T/F_T$ ratio exhibit large values of the average Gini index,
indicating an unequal distribution of females among males. In contrast, regions with a smaller $M_T/F_T$ show a rapid decrease in the Gini value, although it never reaches 0, meaning that the females never distribute equally between the total number of males, even though all the males are equally effective and the all territories offer the same amount of resource.

This last result corresponds to the Gini index calculated over all males, regardless of whether they are associated with females—a measure we will refer to as the \textit{Global Gini index}. However, the values may be strongly influenced by bachelors, who are not linked to any harem. For this reason, it is useful to also consider the Gini index excluding these individuals from the system, which we will call the \textit{Group Gini index}. 

\begin{figure}[b]
\centering
\includegraphics[width=\columnwidth]{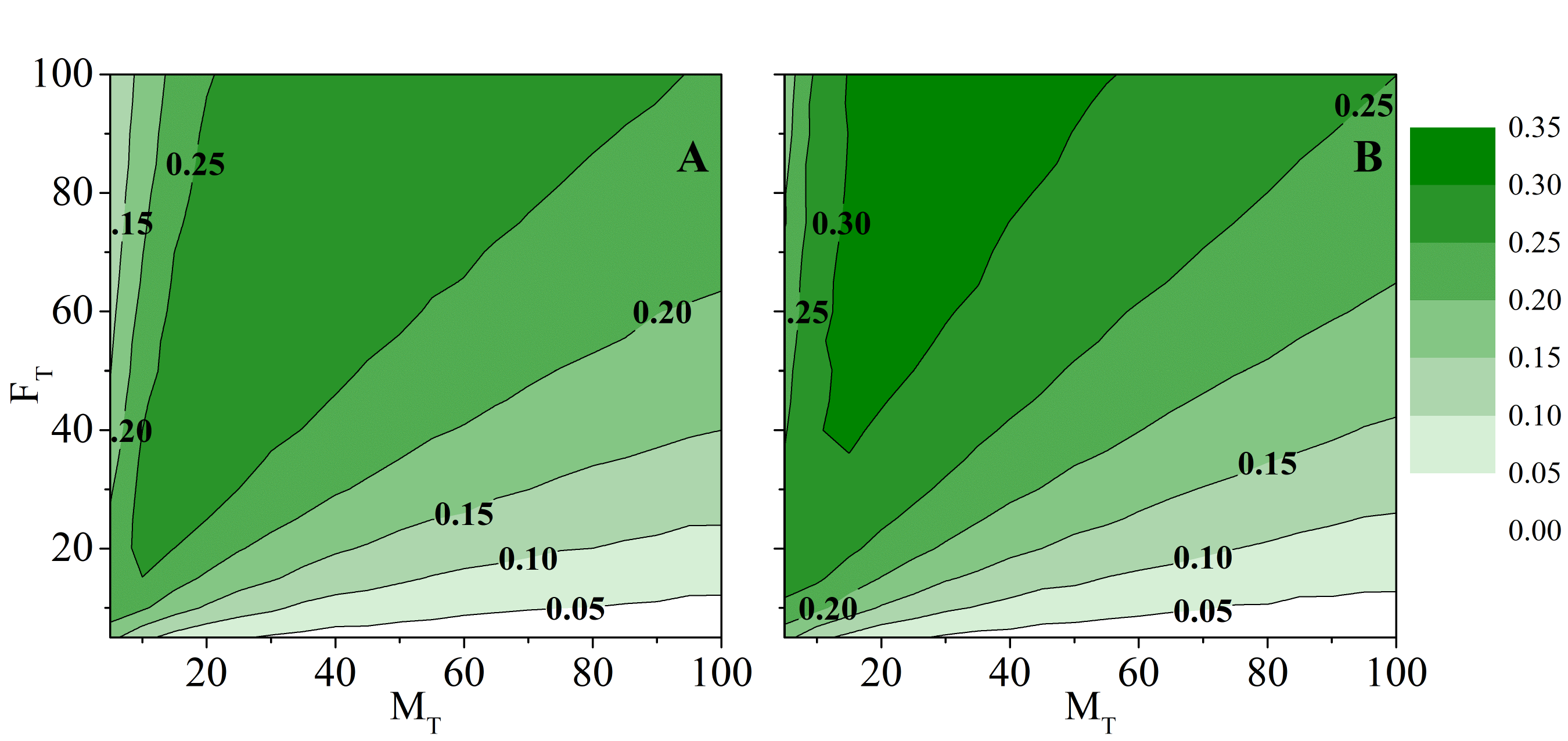}
\caption{Heat maps of the Group Gini index (excluding bachelor males) for different combinations of total number of males ($M_t$) and females ($F_t$) in the system, for two different levels of resources, $R$. A) $R = 20$; B) $R = 500$. Each pixel is the average of 20 realizations. Other parameters are: $H = 5$, $c = 1$, $\beta = 1$, $\sigma= 1$. Note that the color-scale is limited from 0.0 to 0. 35 for visualization purposes.} 
\label{fig:fig3}
\end{figure}

In this case, we observe an unexpected pattern, which is shown in Fig.~\ref{fig:fig3}.

While the Global Gini index exhibits a monotonically decreasing trend as the $M_T/F_T$ ratio decreases (Fig.~\ref{fig:fig2} B), the Group Gini index shows a non-monotonic pattern, with the highest value occurring in a region characterized by a moderate, rather than minimal, number of males and a larger number of females (Fig.~\ref{fig:fig3}).
This result is particularly surprising, considering that, in this simplified version of the model, all males and territories are equally effective, making such an outcome entirely unexpected. As shown Fig.~\ref{fig:fig3} A, increasing resource richness does not significantly alter the overall Gini distribution, but it does raise the maximum level of inequality observed. Table~\ref{tab:table2} presents the mean minimum and maximum Group Gini values corresponding to this same map, comparing systems with limited and abundant resources in both, homogeneous and heterogeneous settings, as be discussed below.

\begin{table}[t]
\caption{Mean maximum and minimum Group Gini and their corresponding standard deviations, for homogeneous and heterogeneous versions of the system under scenarios with limited ($R=20$) and abundant ($R=500$) resources, hereafter referred to as `poor' and `rich' systems, respectively. 
$U(i,j)$ denotes a uniform distribution over the interval from $i$ to $j$ for the corresponding parameter.}
\label{tab:table2}
\centering
\begin{tabular}{cccc}
$R_{i}$ & $\sigma_i$ & $G_{max}$  $(SD)$ & $G_{min}$  $(SD)$ \\ \hline
20 & 1 & 0.289 (0.003)& 0.016 (0.007)\\ 
500 & 1 & 0.321 (0.006) & 0.013 (0.007)\\ 
{$\in$ U(1,39)} & {$\in$ U(0,1)} & 0.323 (0.009) & 0.024 (0.007)\\ 
{$\in$ U(1,999)}  & {$\in$ U(0,1)} & 0.391 (0.015) & 0.024 (0.006)\\ 
\end{tabular}
\end{table}

A deeper analysis of the system reveals additional features in this same region of the map that help explain the higher values of the Group Gini index. These features are related to what we refer to as \textit{environmental pressure}.

As defined in our payoff function (Eq.~\ref{eq:2}), the total cost that females must face is composed of a linear term with an intercept ($H$), representing pressure from external factors, and a variable term that captures the social pressure exerted by bachelors ($\beta B$). This combined linear structure, which encompasses both external and social influences, defines what we call environmental pressure.

\begin{figure}[h]
\centering
\includegraphics[width=\columnwidth]{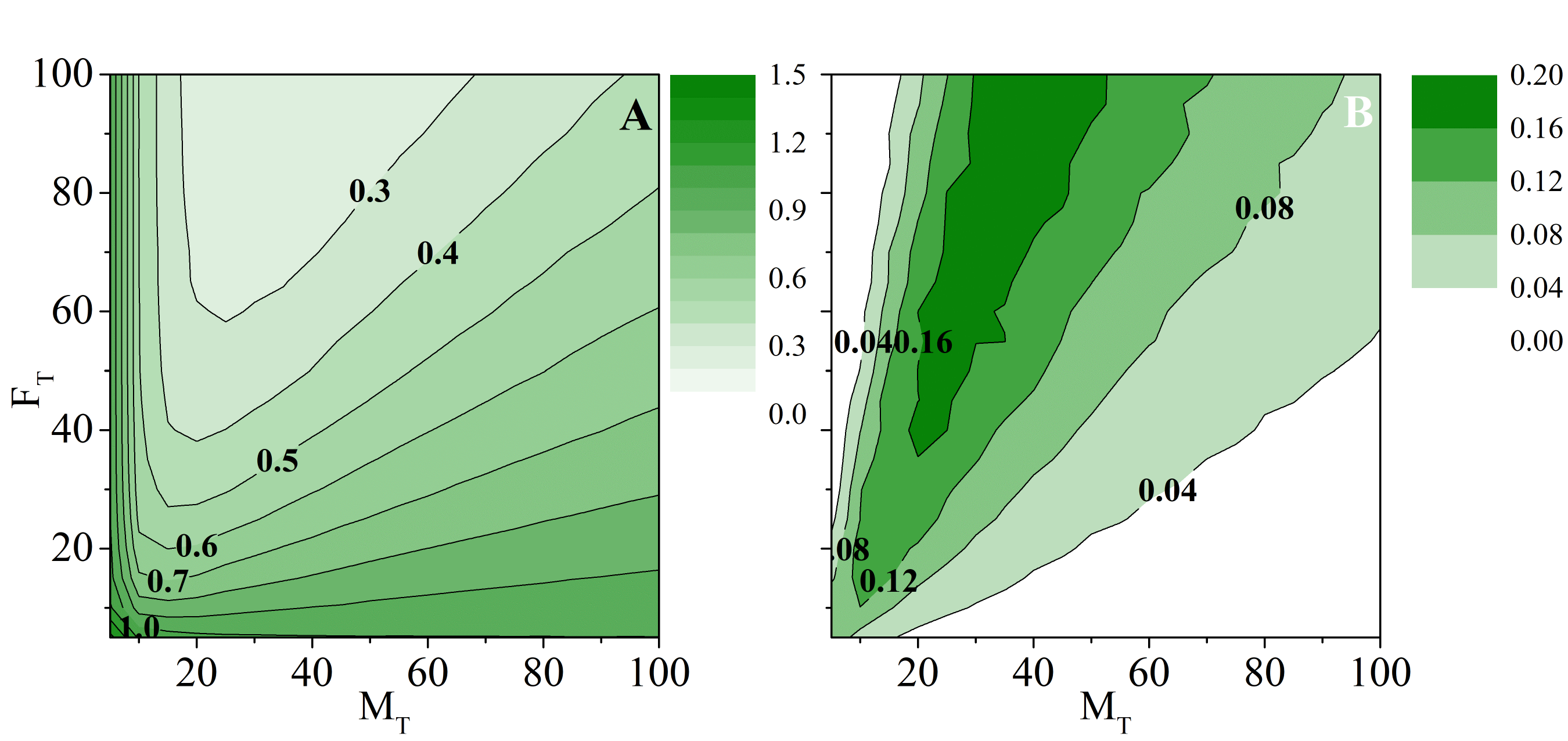}
\caption{A) Heat map of the average environmental pressure over the number of males in the system. B) Coefficient of variation of environmental pressure for different combinations of number of males ($M_t$) and females ($F_t$) on the system, for two different levels of resources ($R$). Each pixel is the average of 20 realizations. ($R_i = 20~\forall i$, $H = 5$, $c = 1$, $\beta$ = 1, $\sigma_i = 1~\forall i$.) }
\label{fig:figV}
\end{figure}

In Fig.~\ref{fig:figV}A, we present the average environmental pressure normalized by the total number of males ($M_T$), given by:
\[
\frac{H + \beta B}{M_T},
\]
As observed, the lowest values occur in the same region where the Group Gini index reaches its highest levels, an outcome that can be directly attributed to the larger number of familiar groups, as shown in Fig.~\ref{fig:fig2}.

Furthermore, Fig.~\ref{fig:figV}B presents the coefficient of variation (defined as the ratio of the standard deviation to the mean) for the normalized environmental pressure. Interestingly, this same region exhibits the greatest temporal variability. This suggests that, although the average environmental pressure on females is lowest in this region, its fluctuations over time are more pronounced than elsewhere on the map. In subsequent analysis, we observed that, in this region of the map, the bachelors in the system (which represents a minority of the total males, as shown in Fig.~\ref{fig:fig2}A) mostly constitute favorable options for the females, offering a positive potential payoff offer, which explains the higher rates of temporal variability.

These two findings help explain why females may distribute more unequally. When environmental pressure increases due to the temporary presence of bachelors, some females—possibly those in excess relative to the number of males—are forced into larger groups. When the pressure subsides, the associated reduction in costs allows these large groups to persist, even if the benefits (first term of Eq.~\ref{eq:2}) are comparatively smaller. As a result, the recurring presence of bachelors may further contribute to the long-term stability of these unequal distributions.

Finally, in a rich system, this reduction in environmental pressure is less marked, leading to higher minimum values on the map. This may account for the even higher Gini values observed in that region when resources are abundant.

\subsection*{Heterogeneous systems}
We call heterogeneous those systems where territories have unequal resource availability ($R_i \neq R_j, \forall i \neq j$) and males exhibit varying effectiveness ($\sigma_i \neq \sigma_j, \forall i \neq j$). We analyzed simple scenarios, where both quantities are random, with uniform distribution $U$ between fixed bounds. For $\sigma_i$ we used:
\[
    \sigma_i  \in U(0,1),
\]
while for $R$ we used: 
\[
R_i  \in U(1,b).  
\]
In the following, we refer to systems with $b = 39$ as `poor', and those with  $b = 999$ as `rich'. The value of these upper bounds were chosen based on the expected mean value for a uniform distribution:
\[
\mathbb{E}[R] = \frac{a + b}{2},  
\]
so these systems were comparable to their homogeneous equivalents discussed above.

We obtained heat maps for the proportion of males with a harem, and a the Global Gini index, that are almost identical to that of the corresponding homogeneous systems (Fig.~\ref{fig:fig2}), indicating that heterogeneity does not significantly impact these population features, indicating that heterogeneity does not have a significant qualitative impact on these population features. We also observed similar patterns in the distribution of the Gini index for the set of harems. However, male and territory heterogeneity appears to increase the maximum value of this index, amplifying the effect of resource availability noted in the previous section (see the last two rows of Table~\ref{tab:table2}).

\subsection{Stressful scenarios}
Another set of simulations was performed considering larger
values of $H$, which represents the mean field costs or stress (this could be interpreted as larger predation risk, inter-specific competition, or hostile climate). Even under these conditions, the previous results remained robust, showing the same global qualitative patterns observed in the previous figures.

\subsubsection{Group switch penalty}

After incorporating the penalty term $\eta$ into each payoff distribution to discourage group switching, we did not observe notable changes in the heat maps shown in Fig.~\ref{fig:fig2} or in the stability of the global distribution previously evaluated with the Kolmogorov–Smirnov method.

However, when analyzing the Group Gini Index (as in Fig.~\ref{fig:fig3}), we detected a significant effect of the penalty term, particularly in homogeneous systems with limited resources ($R_i = 20~\forall i$).

The penalty parameter exhibits a marked interaction with population size and sex ratio $\frac{M_T}{F_T}$, but only when $\eta$ is sufficiently large. To quantify this effect, we computed the \textit{Relative Increment of the Group Gini Index} (RIGGI) as a function of $\eta$:

\[
RIGGI(\eta_i) = \frac{Group~Gini~Index(\eta = \eta_i)}{Group~Gini~Index(\eta = 0)}
\]

\begin{figure}[h]
\centering
\includegraphics[width=\columnwidth]{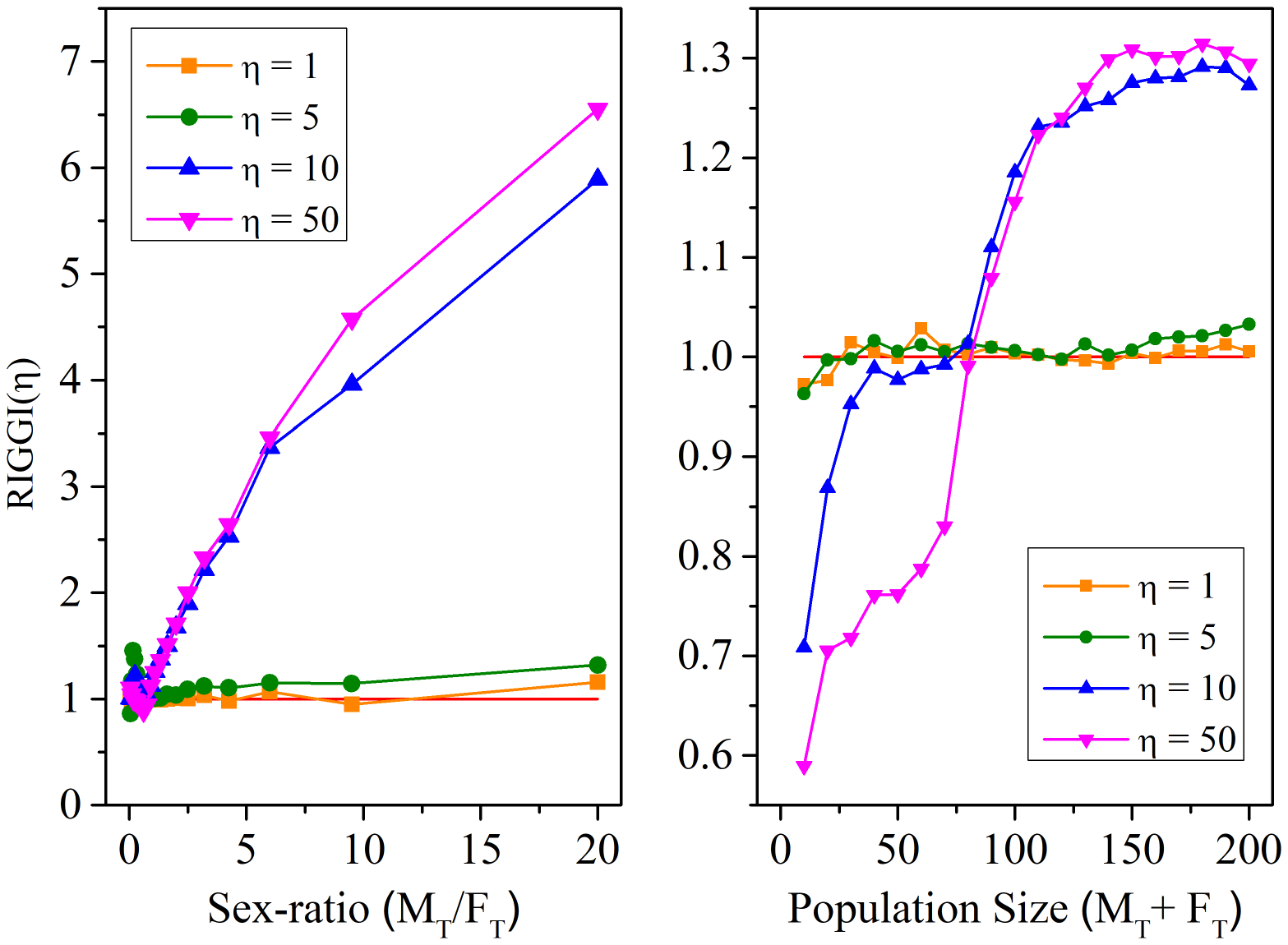}
\caption{(A) Relative increment of the Group Gini Index (RIGGI) as a function of sex ratio ($M_T/F_T$), for a population of 100 individuals. (B) Relative increment of the Group Gini Index as a function of total population size ($M_T + F_T$), assuming a 1:1 sex ratio. The red line on both plots represent a null increment of the Gini Index (RIGGI = 1).}
\label{fig:figC}
\end{figure}

As illustrated in Fig.~\ref{fig:figC}A, when the sex ratio exceeds 1:1, increasing $\eta$ produces an almost linear rise in RIGGI, indicating a higher concentration of females compared to simulations without a switching cost. 
Figure~\ref{fig:figC}B shows a similar, though less pronounced and non-linear, effect associated with total population size under a balanced sex ratio (1:1).

\subsection{Group size strategies}
\begin{figure}[h]
\centering
\includegraphics[width=\columnwidth]{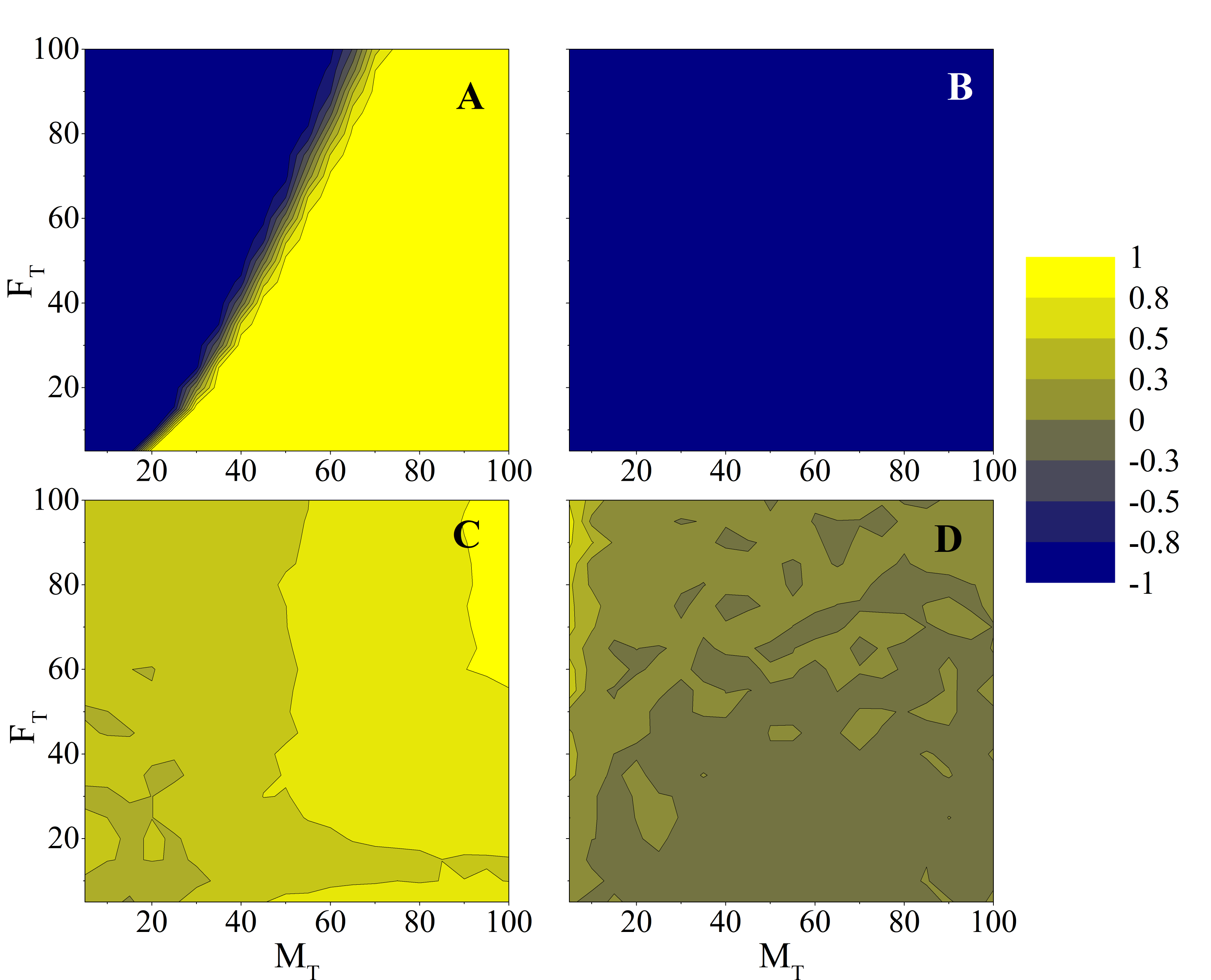}
\caption{Heat maps of Spearman's correlation coefficient between group size and payoff for different combinations of males ($M_t$) and females ($F_t$) in the system. (A) Poor homogeneous system ($R_i = 20,~\sigma_i = 1~\forall i$), statistically significant ($p$-value $< 0.05$). (B) Rich homogeneous system ($R_i = 500,~\sigma_i = 1~\forall i$), statistically significant ($p$-value $<0.05$). (C) Poor heterogeneous system ($R_i \sim U(1,20)$, $\sigma_i \sim U(0,1) $), statistically significant ($p$-value $< 0.05$). (D) Rich heterogeneous system ($R_i \sim U(1,500)$, $\sigma_i \sim U(0,1) $), not statistically significant. Each pixel is the mean of 20 realizations.  }
\label{fig:figQ}
\end{figure}

To better understand the dynamics of group formation, we aim to identify the conditions under which it becomes advantageous for a female to belong to a larger group.
To this end, we analyzed each time step of the simulations, recording both the size and payoff distribution of all groups, in order to assess whether a significant correlation exists between these two aspects.
We evaluated this relationship using Spearman's rank correlation coefficient, which is defined as:

 \begin{equation}
    r_s =1 -\frac{ 6\sum_{i}d_i^2}{n(n^2-1)},
    \label{eq:scc}
\end{equation}
where $d_i$ is the difference between the two ranks of each observation, and $n$ the number of observations. This coefficient has the advantage of being easy to interpret, as its absolute value indicates the strength of the correlation, while its sign reveals whether an increase in one variable corresponds to a rise or decline in the other. Additionally, by considering the range of variable values, it is useful for detecting correlations that are not strictly linear, and based on our previous results with female distribution shown in Fig.~\ref{fig:fig3}, we consider that this property makes this coefficient more adequate \citep{ali2022spearman}.
The results of this correlation analysis are presented in Fig.~\ref{fig:figQ}. 

In Fig.~\ref{fig:figQ} A, we observe that in a poor and homogeneous system, two distinct regions emerge, separated by a clear boundary. On the left, where the $F_T/M_T$ ratio exceeds 1, the correlation coefficient is nearly -1, indicating that larger groups always have a lower payoff. On the right, where $F_T/M_T$ is lower, larger groups tend to have a higher payoff. It is useful to analyze this map in conjunction with Fig.~\ref{fig:fig2}, as the region with negative correlation corresponds to where most males have a harem, while the positive correlation region is where a moderate to low proportion of males have a harem. This pattern makes sense, as the benefit of being in a larger group increases with environmental pressure (including the bachelors). We will revisit this topic in the next section.

In Fig.~\ref{fig:figQ} B, we observe that larger groups are consistently at a disadvantage compared to smaller ones in a rich homogeneous system. In contrast, Fig.~\ref{fig:figQ} C, representing a poor and heterogeneous system, shows the opposite pattern: larger groups are favored, especially when the number of males and females in the system increases.

Finally, in the rich heterogeneous system, the correlation between group size and payoff becomes not only weaker but statistically insignificant, as seen in Fig.~\ref{fig:figQ} D.

\section{Mean field analysis}
An analytical insight into the properties of our model can be obtained by examining the partial derivatives of the payoff defined by Eq.~(\ref{eq:2}):
\begin{align}
\frac{\partial Q(F_i,B)}{\partial B} &= -\frac{\beta}{1+F_i},\label{eq:4}\\
\frac{\partial Q(F_i,B)}{\partial F_i} &= \frac{H +\beta B}{(1 + F_i)^2} - \frac{c \sigma_i R_i}{(1 + cF_i)^2}. \label{eq:3}
\end{align}
These two functions express how the payoff of a harem responds to an increase in the number of bachelors in the system, or to an increase of the numbers of females in the group. For clearer interpretation, it is important to highlight that the current payoff and its change reflect the potential attractiveness of the groups for new females.

Equation~(\ref{eq:4}) is always negative, implying that the potential payoff (and future attractiveness for new members) of any harem decreases with an increase in the number of bachelors in the system. However, this decline can be reduced by an increase in the number of females in the harem.

On the other hand, Eq.~(\ref{eq:3}), which shows how attractiveness changes based on the number of females, reveals a more complex behavior, that we analyze below.

\begin{figure}[h]
\centering
\includegraphics[width=\columnwidth]{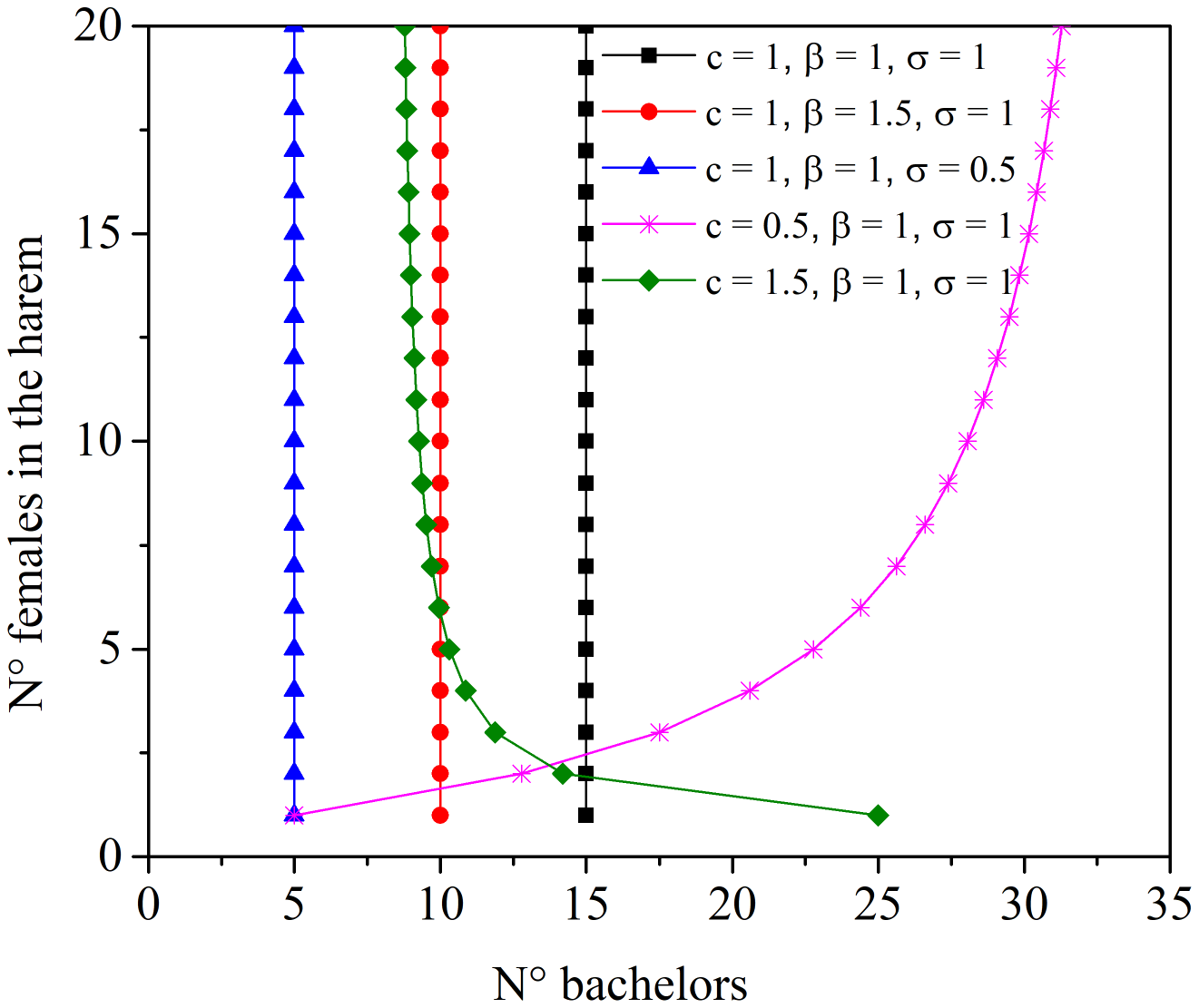}
\caption{Females-Bachelors plane: The threshold between positive and negative derivatives exhibits distinct behavior depending on the females' consumption rate ($c$), as shown in the figure ($R = 20$, $H = 5$).}
\label{fig:figZ}
\end{figure}

In Fig.~\ref{fig:figZ} we show the female-bachelor phase plane of the system. The lines represent, for different sets of parameters, the sign change of Eq.~(\ref{eq:3}).  For example, when the parameter $c$ equals 1 (indicating that each female has the same resource consumption as the male), this frontier is a vertical line, parallel to the females axis,  indicating that the change of behavior of the payoff is independent of the group size. Three such cases are shown, for different values of the other parameters. To the right of this threshold, Eq.~(\ref{eq:3}) is positive, meaning that adding new females to the harem increases the potential payoff of the group, making it more attractive for additional females. 

A different situation is seen when $c$ is larger than 1, i.e. females consume more resources than males, this threshold becomes a non-linear concave function, and the bachelor-threshold value decreases as the number of females in the harem grows until it reaches an asymptotic value. Conversely, if females consume fewer resources than males, the threshold is also concave, but in the opposite direction, meaning the bachelor-threshold value increases as the number of females in the harem increases.

This threshold function can be obtained analytically from  Eq.~(\ref{eq:3}) by solving for $B$. When the parameter $c=1$, the threshold is:
\begin{align}
B_i^* =  \frac{\sigma_i R_i - H}{\beta} .\label{eq:5}
\end{align}
However, if $c \neq 1$, the expression of $B_i^*$ becomes more involved:

\begin{align}
B_i^* = \left(\frac{1+F_i }{1 + c F_i}\right)^2\frac{c R_i \sigma_i - H}{\beta}
.\label{eq:6}
\end{align}

These behaviors might seem counterintuitive. Why does a group with more `gluttonous' females have a smaller threshold? We can understand this by examining Eq.~(\ref{eq:3}), as discussed below.

We can first observe that the positive term of the equation decreases as the number of females in the group ($F_i$) increases. In terms of the real system, this term can be interpreted  in relation to the previously defined environmental pressure. It plays the role of an ``environmental pressure per capita'': as the number of females in the group increases, or the environmental pressure decreases, the per capita pressure declines (or, in other words, the environmental adjustment increases). But, in Eq.~(\ref{eq:3}), a decrease in the per capita environmental pressure is initially a decrease in the derivative itself, meaning that each new female constitutes a smaller contribution to the group, eventually reaching the point when it may not be a contribution at all, or be even become detrimental.

The second term can be decomposed as the product $cZ$, where $Z=\sigma R(1+cF)^{-2}$ represents the effective yield of the territory relative to the group size. Therefore, the product $cZ$ approximates how much the female can actually consume in that group. We will refer to this term as the ``social adjustment.'' If this term is large in relation to the first one---the environmental pressure per capita---, the derivative could be weakly positive or even negative, meaning that adding a new female would have a smaller contribution or be detrimental, respectively. In the context of the system, if the current territory provides a large yield to each female, and the per capita pressure is not high, adding a new female might not be beneficial, on account of the decrement of the yield.  However, in the opposite case, if the environmental pressure per capita is too large and the social adjustment is relatively smaller, the addition of a new female can be favorable, if the contribution to a reduction in the environmental pressure is more significant than the decline in the territory yield and social adjustment. 

With these ideas in mind, it becomes more clear that a group with more resource-demanding females ( $c >$ 1) will experience a worse social adjustment compared to other cases. This occurs because, as each female consumes more, the share of total resources ($R_i$) available to each individual decreases. However, when the number of bachelors in the system rises, the environmental pressure intensifies. The threshold represents the point at which this pressure becomes so large that the addition of a new female offsets the already low social adjustment by improving the group's environmental adjustment. 

For this same reason, we observe that, when the harassment pressure $\beta$ increases, or the male's effectiveness $\sigma$ decreases, the threshold frontier moves to the left in Fig.~\ref{fig:figZ}.

In the opposite case, where females have smaller resource demands, the group already benefits from strong social adjustment, as the territory provides a larger effective yield. This advantage is so substantial that even as the number of bachelors —and consequently, environmental pressure— increases, the group maintains a stable balance. Only when the larger threshold is reached does the addition of a new female become beneficial.

\subsection{Back to the simulations}
\begin{figure}[h]
\centering
\includegraphics[width=\columnwidth]{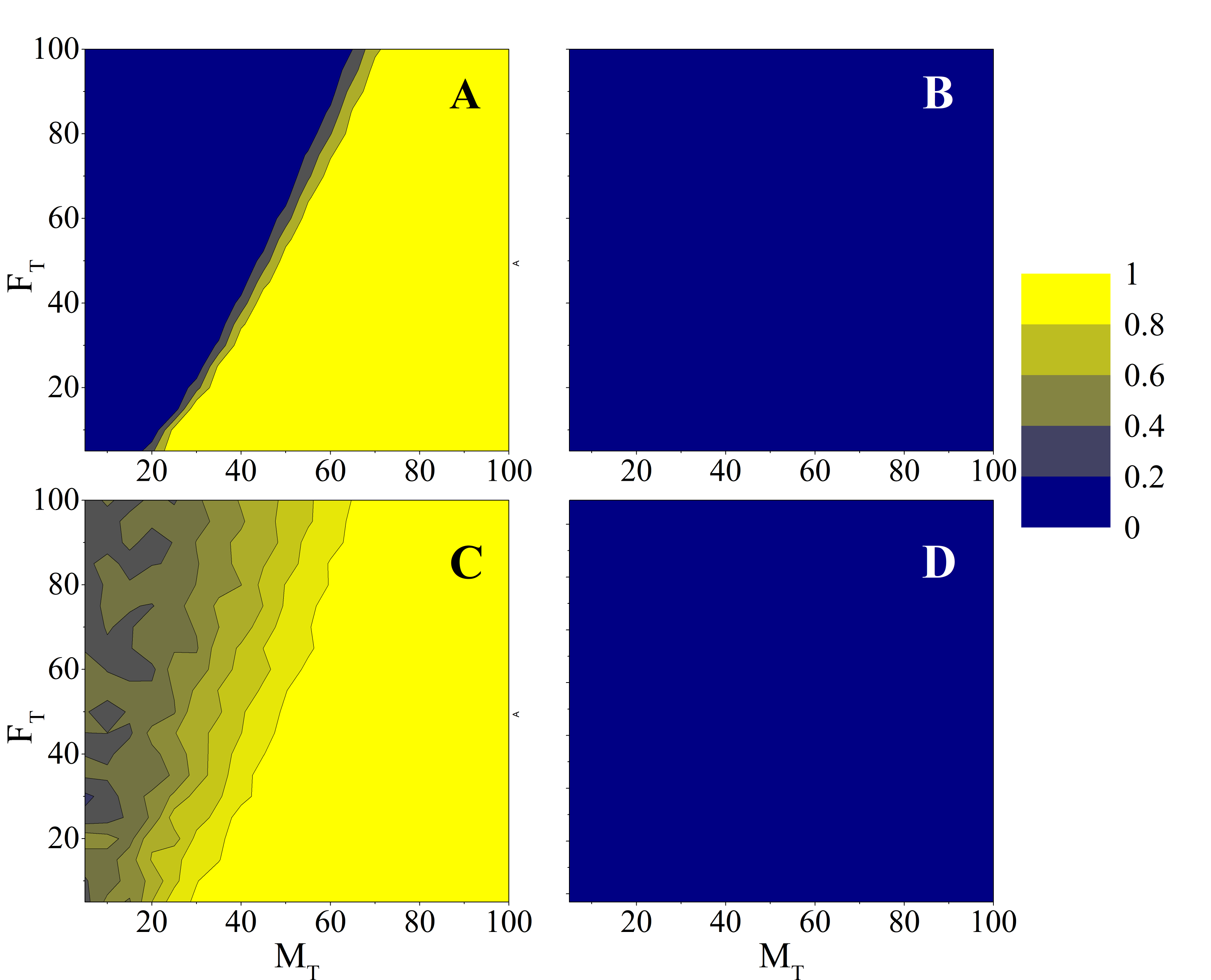}
\caption{Heat-maps of the proportion of familiar groups that reach its particular $B$-threshold, for different combinations of number of males ($M_t$) and females ($F_t$) on the system: A) poor homogeneous system ($R_i = 20~\forall i$, $\sigma_i = 1 ~\forall i$); B) rich homogeneous system ($R_i = 500~\forall i$, $\sigma_i = 1~\forall i$); C) poor heterogeneous system ($R_i \sim  U(1,20)$, $\sigma_i \sim U(0,1) $); D) rich heterogeneous system ($R_i \sim  U(1,500)$, $\sigma_i \sim U(0,1)$). Each pixel is the average of 20 realizations. }
\label{fig:figX}
\end{figure}

We performed an additional set of simulations, focusing on the simpler case of $c = 1$, to register the proportion of familiar groups that reach its particular $B$-threshold ($B_i^*$) in each scenario. The results are illustrated in Fig.~\ref{fig:figX}.
We observe that the same border noticed in Fig.~\ref{fig:figQ}A is present at Fig.~\ref{fig:figX}A, which also shows a total correspondence between the scenarios where the familiar groups surpass their thresholds and when the larger group size is more beneficial. However, in additional simulations, we observed that at a higher set of cost values (i.e. , $H$ and $\beta$), this border between regions becomes steeper, with a reduction of the blue region (negative correlation), which finally disappears when the total costs are larger than the offer (which is $R_i = 20~\forall i$). 

This same association can be observed in Fig.~\ref{fig:figX}B, which shows that in a rich homogeneous system, no groups reaches the threshold, and having a larger group gives the lowest payoff. 
In the heterogeneous system, we do not find this clear association with the B-threshold, as we observe that the patterns on the maps of Fig.~\ref{fig:figX} do not correspond to the ones of the Fig.~\ref{fig:figQ}, unlike the homogeneous system.

After further analysis and simulation, we observed that the patterns of correlation between payoff and groups size, are almost the opposite of the maps for correlation between payoff and group's intrinsic quality ($R_i \sigma_i$).

\begin{figure}[h]
\centering
\includegraphics[width=\columnwidth]{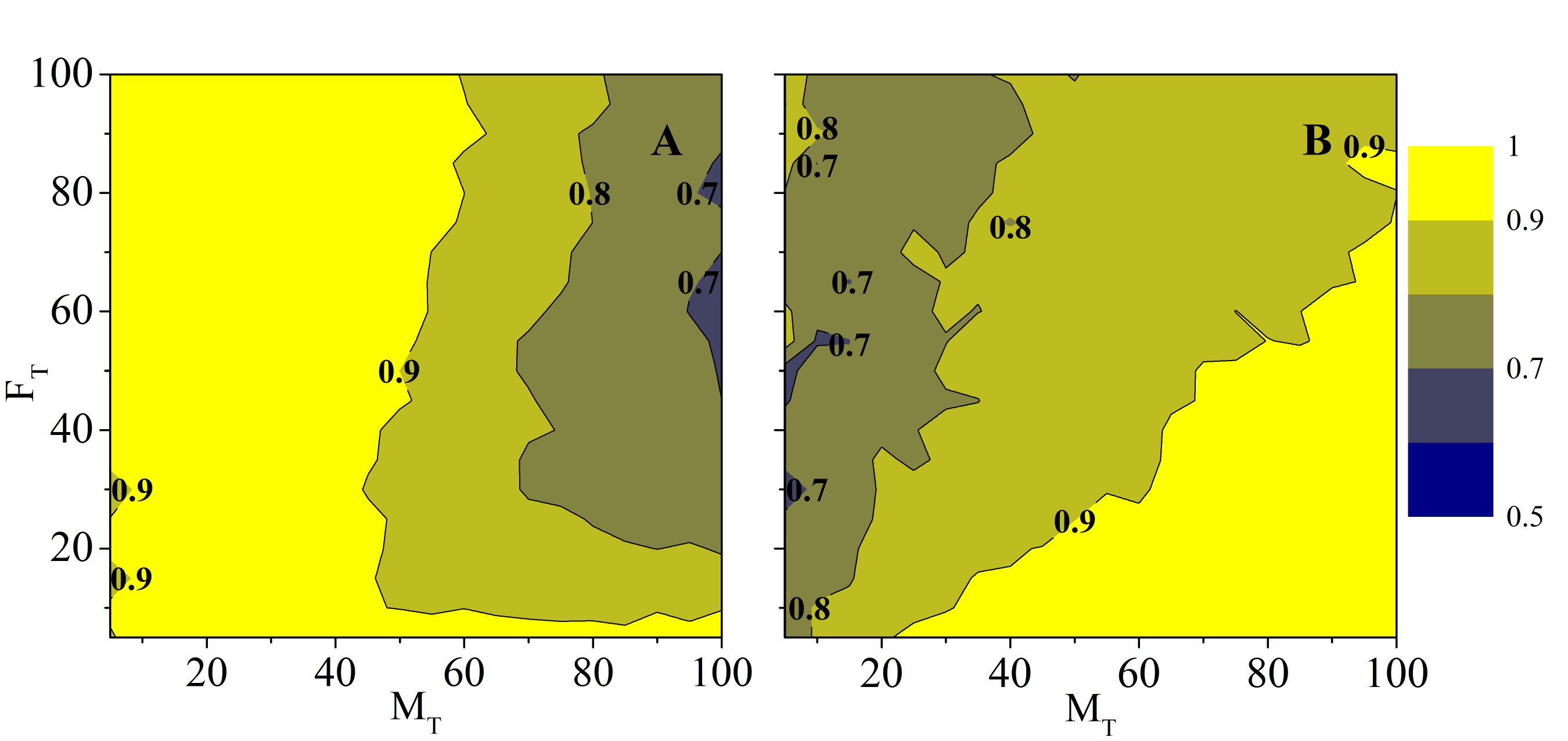}
\caption{Heat maps of Spearman's correlation coefficient between group intrinsic quality and payoff for different combinations of males ($M_t$) and females ($F_t$) in the system: (A) poor homogeneous system ($R_i = 20~\forall i$, $\sigma_i = 1~\forall i $), statistically significant ($p$-value $< 0.05$); (B) rich homogeneous system ($R_i = 500 ~\forall i$, $\sigma_i = 1~\forall i $), statistically significant ($p$-value $< 0.05$). Each pixel is the average of 20 realizations.}
\label{fig:figW}
\end{figure}

In Fig.~\ref{fig:figW}A, we can observe that, in the regions where the payoff correlates more strongly with intrinsic quality, the correlation with group size decreases and vice versa. This would explain why the patterns of group size-payoff correlation do not correspond completely with the $B$-threshold map because the number of individuals of each group is overshadowed by the effect of system heterogeneity.

This could explain why we observed that, in Fig.~\ref{fig:figX}D, almost no group reaches the threshold, but, unlike Fig.~\ref{fig:figQ}B, the system loses any significant correlation between group size and payoff, while the correlation between group intrinsic quality and payoff is always high in the same system (Fig.~\ref{fig:figW}B).

\begin{figure}[h]
    \centering
    \includegraphics[width=\columnwidth]{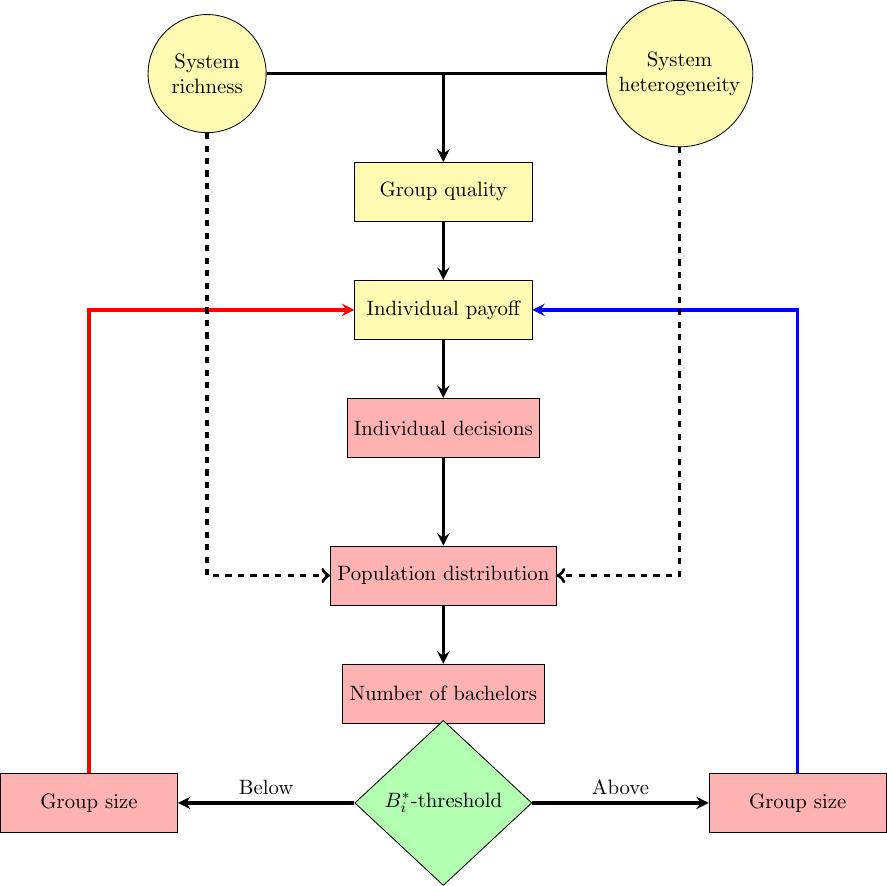}
    \caption{Summary of the current results. Yellow nodes indicate aspects that are deterministic (by chosen parameters or deterministic functions); red nodes indicate aspects that are are stochastic or emergent; green nodes indicate aspects that are analytical (from mean field). The red arrows indicate negative correlation; the blue arrows indicate positive correlation.}
    \label{fig:flowchart}
\end{figure}

\section{Discussion}

According to optimal foraging theory, herbivores are expected to adopt adaptive strategies that maximize energy intake while simultaneously reducing associated risks \citep{sih1980optimal}. Camelids, like other ungulates, develop complex social grouping behaviors that should provide greater benefits to individuals compared to solitary living \citep{iranzo2018predator, taraborelli2012cooperative, creel2014effects}.

However, the literature indicates that both male and female camelids face different costs and trade-offs associated with group living, with each individual pursuing strategies that maximize its own fitness. For instance, variation in harem size reflects a set of discrete social strategies, each associated with different fitness outcomes for males and females  \citep{cassini2009sociality, sutherland1996individual,book}.

These features could also affect the way in which an emergent event, such as the spread of an infectious agent, takes course in the population, or the survival rate during a period of harsh weather. Moreover, such events could alter the age and sex structure and the dynamics of the population, thereby influencing the future allocation of females among groups \citep{puig2000dinamica, gonzalez2022historical, ferreyra2022sarcoptic, gonzalez2024mean, mysterud2002role}. 

Modeling is a valuable tool for detecting and analyzing emergent properties that arise at higher levels of organization and are not directly deducible from observations at the individual level \citep{breckling2005emergent,nielsen2000emergent}.  
Among the different modeling approaches, agent-based and individual-based models have proven highly effective for developing and testing ecological theory. They help elucidate how interactions among individuals shape the dynamics of ecological systems and determine the system’s global properties. Additionally, they allow us to explore how such systems might respond to changes in environmental conditions \citep{dumont2004spatially, mclane2011role, he2022agent, vodopivec2025spatially}.

In our model, we define a behavioral rule for each female---expressed as a mathematical function---based on ecological knowledge of the species. We then observe how the collective implementation of this rule gives rise to unexpected patterns in population organization and dynamics.

Since the Gini index is not commonly used in ecological studies, the interpretation of its absolute values is not straightforward. To provide an empirical reference, we estimated the expected range of the Group Gini index using field-based group size data reported for wild camelid populations (see Appendix~A) \citep{puig2007distribucion, marino2012indirect, marino2014ecological, bonacic2002density, karandikar2023spatial}.

Interestingly, the values obtained in our model simulations fall within the same range ($\sim 0.2$–$0.4$), suggesting that the level of inequality emerging from the model is consistent with observed population structures. This agreement supports the relevance of the proposed individual-based mechanisms and indicates that realistic patterns of group-size inequality can emerge without fine-tuning parameters to empirical distributions. 

Returning to our earlier theoretical discussion of group size, we find that the benefits a female gains by joining a larger group to maximize her payoff are strongly influenced by habitat richness, environmental heterogeneity and the sex ratio. These factors affect group dynamics not only in the additive fashion, but also through complex interactions (Fig.~\ref{fig:figQ}).

The literature suggests the existence of a resource threshold beyond which group living becomes beneficial \citep{mcnab1963bioenergetics, bowyer2020evolution}.
Similarly to this hypothesis, we identified a  threshold value specific to each familial group, which integrates both social and environmental factors, and delineates the region in which the addition of a new female increases the attractiveness of the harem to other females in the system  (${dQ}/{dF} > 0$). This threshold ($B_i^*$, Eq.~(\ref{eq:6})) can be interpreted as the minimum level of environmental pressure on the group---relative to the quality of the occupied territory---beyond which the benefits of incorporating additional females outweigh the associated costs. The probability that a harem exceeds this threshold directly affects the advantage of larger group sizes: only when this condition is met does group size show a positive correlation with individual payoff.

Building on this idea, an environmental pressure threshold could be defined as a function of predator density or predation rate within a similar modeling framework operating at a different temporal scale. Such an outcome would align with field observations in wild camelid populations, where group size has been shown to vary with predator occurrence and perceived predation risk
\citep{iranzo2018predator,taraborelli2014different} 

It has being stated that gregariousness in ungulates is influenced by both the quality and spatial distribution of resources. This aligns with our results, where we observed an increase in the Group Gini Index correlated with increases in both overall resource richness and heterogeneity within the system (Fig.~\ref{fig:fig3}) \citep{bowyer2020evolution,jarman1974social}.
Moreover, it is expected that an even and predictable distribution of resources could promote gregarious behavior in these species. Our model partially supports this hypothesis, but it also introduces the influence of sex ratio, arising from the differing behaviors exhibited by males and females, as shown in Fig.~\ref{fig:figQ}A and B \citep{cassini2009sociality, bowyer2020evolution}.

Our analysis of the effects of incorporating the penalty parameter $\eta$ into the payoff function, which represents the cost of switching groups, revealed a notable interaction with population size and sex ratio. However, this pattern was robust only in homogeneous systems with limited territorial resources. This may be because variability in territory quality can become a stronger determinant of female decision-making than the switching cost itself, while abundant resources may compensate for the added penalty. In contrast, in systems where resources are limited across all territories, larger population sizes (implying more competing females or bachelors) or higher sex ratios (more bachelors per female) can strongly influence female organization. When switching costs are included and not offset by resource abundance, these factors can substantially modify the distribution of females among groups.

Nevertheless, aggregation costs are expected to set an upper limit to the group size of wild ungulates, a limit that was not observed in our model. However, it is likely that such an upper bound could be imposed by an active male, who must also perform an energetic balance involving the demands of vigilance, social monitoring and competition with other males \citep{jarman1974social, lung2007influence, marino2012indirect, marino2014ecological}. In line with this, it has been observed that camelid males regulate their group sizes by rejecting new females, and that population density has only a weak effect on increasing family group size \citep{marino2014ecological, arzamendia2018social}.

Our results also indicate that larger groups are less favorable in homogeneous, resource-rich environments. Even under a greater environmental pressure (such as intensified sexual harassment by bachelors), these costs are diminished by the abundance of resources. In such cases, the addition of a new female primarily results in reduced per capita resource availability within the group, thereby discouraging further aggregation. 

It is also remarkable how the value of strategies, such as group size, is also conditioned by the female distribution and  the number of bachelors, which is not a deterministic but an emergent feature of the population. In other words, we can clearly observe how independent ---female--- individuals modify the population system through their respective decisions, and subsequently the system conditions each individual fitness (Fig.~\ref{fig:flowchart}). 
However, the influence of this threshold becomes less direct---or even entirely obscured---when the system (i.e., territories and males) is heterogeneous. The heat maps suggest that in poor, heterogeneous environments, regions where the correlation between group size and individual payoff is weak tend to show a stronger correlation between group intrinsic quality ($\sigma_i R_i$) and payoff, and vice versa. In rich, heterogeneous systems, this pattern becomes even more pronounced, with group quality potentially replacing group size as the dominant weighting factor in female decision-making.

This aligns with ecological evidence suggesting that habitat structure, as well as forage quality and availability, play a central role in shaping group cohesion and behavior in wild camelids \citep{creel2014effects, ripple2003wolf, moll2016spatial}. Our results suggest that these same ecological factors may directly influence not only the formation of larger social groups, but also the degree to which larger groups provide a fitness benefit. 

Other models that try to analyze the self-organization of individuals into groups, like the ones collected in Gerard et al. (2002), take into account variables such as space, movement, and habitat openness. We consider these factors to be of great importance, as well as others, in social and population dynamics, as suggested by research \citep{iranzo2018predator, cassini2009sociality,smith2020and}. Even so, we perform this reductionist model, taking into consideration only female's choice, to observe the emergent properties at the population level, so we could deep into the understanding of which kind of properties could be expected just by the individual action of females looking to maximize their own fitness. 

Sibly's dynamical model of group formation states that optimal group sizes are inherently unstable due to the independent behavior of individuals seeking to maximize their own fitness \cite{sibly1983optimal, gerard2002herd}. Our results partially support this theory, as we observed group sizes fluctuating over time, even at the cost of reducing the collective payoff. However, we also found that there was not always a significant correlation between group size and payoff/fitness. The results of our stochastic simulations do not suggest the existence of a fixed equilibrium group size,  but rather indicate a stable distribution 
 of the entire system (Fig.~\ref{fig:figK}),  similar to the results from Gerard \& Loisel's model (1995) \citep{gerard1995spontaneous}. However, unlike in their model, where the distribution represents an equilibrium, the stable distribution in our case emerges as a property of the population's individual behaviors. Notably, this final distribution in our model is less sensitive to increases in population density than the equilibrium distribution observed in Gerard and Loisel's model. 

Several studies on sexual selection and polygyny in wild mammals have contributed to the construction of a general ecological theory on the subject, distinguishing, for example, between resource defense polygyny and female defense polygyny. Although we do not propose these categories as strictly separate or mutually exclusive, our simulated system lies somewhere between the two, or perhaps in a transitional scenario, in which males not selected for their resource offering adopt more aggressive behaviors \citep{emlenecology, griffin2019insect}.
In our case, familiar males do not play a defensive role, and the only mechanism by which females can reduce their payoff costs is through group selection.
Numerous studies of social mammals highlight the central role of female choice in mating patterns, and how this aspect can influence the stability of female distribution among males \citep{hirotani1989social, balmford1992correlates, martin2007long, campbell2008relationship}.
In this sense, our model shows that a population shaped solely by individual female choice can still reach a stable global distribution.

The sexual harassment by bachelor males is a factor often overlooked in the literature, although some authors have highlighted its significance---even comparing it with predation risk---not only in camelids species but also in other mammals \citep{clutton1992mate, cassini1999evolution, galimberti2000frequency}. Our findings support this perspective. Bachelors not only contribute to the environmental pressure experienced by individual females, but they also qualitatively alter the receptivity of harems to the incorporation of new females. This effect, in turn, has notable consequences at the population level.

In our model, bachelors are treated as an independent variable within the system's dynamic, despite their real-world tendency to form groups with their own internal dynamic. This simplification is based on evidence showing that the size and composition of bachelor group fluctuates throughout the day and do not consistently respond to predation risk or habitat structure \citep{cassini2009sociality, vila1995spacing, marino2014ecological}. Nonetheless, we aim to incorporate bachelor group dynamics in future developments of the model, to enrich our understanding of their role.

We also believe that a similar model, formulated in terms of other variables such as the presence of predators, could produce an effect similar to the one that harassment pressure produce in the present one. This factor has being recorded in real systems to have a significant influence in group size and habitat use \citep{smith2019habitat,marino2014ecological, creel2005responses}.

Just as bachelors can destabilize harems through harassment ---particularly when the resident male lacks the capacity to protect females---territorial and aggressive behavior by harem-holding males against bachelors could potentially stabilize group structures, a mechanism not included in the current model \citep{lucherini1996aggressive, clutton1992mate, cassini1999evolution}. For instance, males may reduce the effective presence of bachelors as perceived by females, thereby influencing female decision-making.
Additionally, familiar males may prevent female dispersal and migration to other territories through herding behaviors, as documented for vicuñas in Vila (2000) \cite{comportamiento_vicuna}. We recognize that male-male competition introduces its own layer of complexity. In the same way that, in the present work, we isolated female behavior to study its emergent effects at the population level, we plan to develop a complementary model focused on the dynamics and consequences of male behavior. 

It is likely that male behavior, based on the optimization of their own payoff balance, could define an equilibrium at the group level. This was not observed in our current model, which only produces a stable distribution at the population level. Thus, the combination of both male and female choices could lead to a different stable state for the entire system. Furthermore, it is also likely that real females face some cost associated with switching groups or territories, an aspect that should be considered in future models.

Although our model is based on camelid species as a study system, we consider that it could also be applied to other groups. Nevertheless, given the documented complexity of social behavior in ungulates, we recommend restricting the application of this model to ungulates with polygynous territorial behavior. Other systems, such as leks, would require a different type of formulation \citep{clutton1992mate, bowyer2020evolution, balmford1992correlates}.
In addition, in our model based on wild camelids, the costs associated with predation can be included as a linear term ($H$), supported by field observations and by the specific hunting strategy of their main predator, the puma (\textit{Puma concolor}), a stalk and ambush species. However, predators with a cursorial hunting strategy, such as wolves, would require a more complex mathematical treatment \citep{marino2014ecological, martin2007long, marino2008vigilance, thaker2010group}.
Further developments could incorporate other key factors specific to each system, allowing future models to include additional layers of complexity.

Our model allowed us to feed our theoretical background on the complexity of camelids social behavior, observing and understanding the two-way relationship of the individual and population level.  It also shows how emergent and deterministic features affect the global behavior of the system, which in the end determines the individual fitness.

\section{Data availability}
The computational code generated during the current study are available in the Mendeley Data repository, \url{https://doi.org/10.17632/rhyg8nrk55.1}.

\section{Acknowledgements}
This research was supported by Agencia Nacional de Promoción Científica y Tecnológica (PICT 2019-02167) and Consejo Nacional de Investigaciones Científicas y Técnicas (CONICET, PIP 112-2022-0100160 CO). The authors thank Adrian Monjeau for valuable discussions.

\appendix
\section{Estimation of the Group Gini index from field data}

In this appendix, we provide an empirical reference for the Group Gini index by estimating its expected range from field-based group size data reported in the literature. Since the Gini index is not commonly used in ecological studies, a direct interpretation of its values is not straightforward. To address this issue, we used published group size statistics to generate synthetic distributions and compute the corresponding Group Gini indices for comparison with our model results.

As a validation procedure, we consider several field studies on wild camelid populations in which the mean group size $\mu$ and its standard deviation $\sigma$ are reported. For each study, we generate one thousand synthetic group size samples assuming a normal distribution,
\begin{equation}
x_i \sim \mathcal{N}(\mu,\sigma^2), \qquad i = 1,\dots,1000,
\end{equation}
and compute the associated Group Gini index for each realization.

From these simulations, we obtain an estimated range for the Group Gini index in natural camelid populations between $0.2$ and $0.4$ \citep{puig2007distribucion, marino2012indirect, marino2014ecological, bonacic2002density, karandikar2023spatial}. A selection of field studies used for this estimation, together with the corresponding parameters and resulting Gini values, are summarized in Table~\ref{tab:tableE}.

\begin{table}[h]
\caption{Field records of family group size in wild camelid populations used for the simulations. The table reports the mean group size ($\mu$), standard deviation ($\sigma$), and the Group Gini index obtained from the synthetic distributions.}
\label{tab:tableE}
\centering
\begin{tabular}{cccccc}
Year & Species & $\mu$ & $\sigma$ & Source & Gini \\ \hline
1978 & Vicuña  & 7.38 & 4.05 & Puig \& Videla (2007) & 0.30 \\
1993 & Vicuña  & 6.19 & 2.70 & Puig \& Videla (2007) & 0.24 \\
1978 & Guanaco & 8.14 & 4.26 & Puig \& Videla (2007) & 0.29 \\
1993 & Guanaco & 8.05 & 5.07 & Puig \& Videla (2007) & 0.34 \\
2008 & Guanaco & 6.10 & 2.40 & Marino \& Baldi (2014) & 0.22 \\
2011 & Guanaco & 7.80 & 5.33 & Marino (2012) & 0.38 \\
\end{tabular}
\end{table}

Interestingly, this empirical range coincides with the values obtained in our model simulations. This agreement supports the relevance of the proposed modeling framework and highlights how such approaches can contribute to ecological theory by linking individual-level mechanisms with population-level inequality measures derived from field observations.

\end{document}